\documentclass[12pt,preprint]{aastex}

\input psfig.tex

\begin{document}

\title{Distributions of the Baryon Fraction on Large Scales in the Universe}

\author{Ping He$^{1,2}$, Long-Long Feng$^{3,1}$, and Li-Zhi Fang$^{2}$}

\altaffiltext{1}{National Astronomical Observatories, Chinese
Academy of Sciences, 20A Datun Road, Chaoyang, Beijing 100012,
China}

\altaffiltext{2}{Department of Physics, University of Arizona,
Tucson, AZ 85721}

\altaffiltext{3}{Purple Mountain Observatory, Nanjing 210008,
China}

\begin{abstract}

The nonlinear evolution of a system consisting of collisional
baryons and collisionless dark matter is generally characterized
by strong shocks and discontinuities in the baryon fluid. The
baryons slow down significantly at postshock areas of
gravitational strong shocks, which can occur in high overdense as
well as low overdense regions. On the other hand, the shocks do
not affect the collapse of the dark matter. Consequently, the
baryon fraction would be nonuniform on large scales. We studied
these phenomena with simulation samples produced by the weighted
essentially nonoscillatory (WENO) hybrid cosmological
hydrodynamic/$N$-body code, which is effective at capturing shocks
and complex structures with high precision. We find that the
baryon fraction in high mass density regions is lower on average
than the cosmic baryon fraction, and many baryons accumulate in
the regions with moderate mass density to form a high baryon
fraction phase (HBFP). In dense regions with
$\rho/\bar{\rho}>100$, which are the possible hosts for galaxy
clusters, the baryon fraction can be lower than the cosmic baryon
fraction by about 10\%--20\% at $z\simeq 0$. We also find that at
$z<2$, almost all the HBFP gas locates in the regions with mass
density $0.5 < \rho/\bar{\rho}<5$ and temperature $T>10^5$ K, and
conversely, almost all the gas in the areas of $0.5 <
\rho/\bar{\rho}<5$ and with temperature $T>10^5$ K has high baryon
fraction. Our simulation samples show that about 3\% of the cosmic
baryon budget was hidden in the HBFP at redshift $z=3$, while this
percentage increases to about 14\% at the present day. The gas in
the HBFP cannot be detected either by Ly$\alpha$ forests of QSO
absorption spectra or by soft X-ray background. That is, the HBFP
would be missed in the baryon budget given by current
observations.

\end{abstract}


\section{Introduction}

Although the universe is dominated by dark matter and dark energy,
the observed luminous universe exists in the form of baryonic
matter. The primordial nucleosynthesis predicts $0.015 < \Omega_b
h^{2} < 0.021$ (Walker et al. 1991; Esposito et al. 2000) The best
fitting of cosmological parameters with the temperature
fluctuations of the cosmic microwave background (CMB) radiation
and large-scale structure clustering shows that the mass density
of baryonic mater is $\Omega_b=0.0224 \pm 0.0009$ $h^{-2}$ and the
total matter density $\Omega_m=0.135^{+0.008}_{-0.009}$ $h^{-2}$
(Bennett et al. 2003). Therefore, the cosmic baryon fraction is
$f_{c} \equiv\Omega_b/\Omega_m= 0.166^{+0.012}_{-0.013}$.

In the linear evolution of gravitational clustering perturbations,
the density and velocity distributions of the baryonic gas (or
intergalactic medium [IGM]) are the same as those of the dark
matter field point-by-point on scales larger than the Jeans
length. Even if the IGM is initially distributed differently from
the dark matter, a linear growth mode will lead to the same
distribution of the IGM as of dark matter (Bi et al. 1992; Fang et
al. 1993; Nusser 2000; Nusser \& Haehnelt 1999). Thus, in the
linear regime, although the density distributions of both IGM and
dark matter are inhomogeneous, the baryon fraction is uniform on
scales larger than the Jeans length. However, observations show
that the distribution of the baryon fraction probably is not
uniform. X-ray measurements have revealed that the baryon fraction
in galaxy clusters is less than the prediction of primordial
nucleosynthesis (Ettori \& Fabian 1999). This discrepancy is more
serious in the cores of clusters (Sand et al. 2003). Even
considering a depletion of baryons at the virial radius, the
baryon fraction in galaxy clusters is still less than the
predicted value (Ettori 2003). This indicates that in the
nonlinear evolved fields, the baryon fraction distribution
$f_b({\bf x})$ is spatially dependent, not equal to $f_c$
everywhere.

In this paper, we explore the formation and evolution for the
nonuniform distribution of baryon fraction. On large scales, the
nonuniformity of the baryon fraction is a result of the
statistical discrepancy of baryonic gas from the underlying dark
matter during nonlinear evolution. Recently, the statistical
decoupling between baryonic gas and dark matter fields has been
studied, both theoretically and numerically (Feng et al. 2003; He
et al. 2004; Pando et al. 2004). These studies found that,
although in both the linear and the nonlinear regimes the
evolution of the IGM is dynamically governed by the gravity of the
underlying dark matter field, the effects of the gravity in
different regimes are very different. In the former, the gravity
of the dark matter ensures that baryonic gas follows dark matter
point-by-point, while in the latter, the gravity of the dark
matter will inevitably lead to the decoupling of the baryonic gas
from the dark matter on scales larger than the Jeans length. That
is to say, the linear dynamical behavior of the baryonic matter
can be simply obtained from the dark matter field via a similarity
mapping (e.g., Kaiser 1986), while in the nonlinear regime, the
similarity is broken. Obviously, a direct consequence of the
baryon -- dark matter discrepancy is the inhomogeneity of the
baryon fraction distribution. We follow this clue to study the
properties of the deviation of $f_b({\bf x})$ from $f_c$.

The outline of this paper is as follows. In \S 2 we describe the
dynamical mechanism leading to the deviation of the baryon
fraction from the cosmic value, together with the predictions from
this mechanism. In \S 3 we present the samples used to numerically
study the baryon fraction. In \S 4 we investigate the statistical
features of the field $f_b({\bf x})$, and compare them with the
predictions. Finally, conclusions and discussions are given in \S
5.

\section{The mechanism leading to non-uniform distribution of
  baryon fraction}

A dynamical mechanism of separating baryonic gas from dark matter
was addressed in the early study of structure formation (Shandarin
\& Zel'dovich 1989). Because the dark matter particles are
collisionless, the velocities of the dark matter particles are
multivalued at the intersection of the dark matter particle
trajectories. On the other hand, the IGM, as an ideal fluid, has a
single-value velocity field. Thus, discontinuities, such as shocks
or complex structures, will develop in the density and velocity
fields of gas at the intersection of the dark matter particle
trajectories. This leads to the decoupling between the mass and
velocity fields of the IGM and the dark matter. This feature can
also be seen with the self-similar solution of spherical collapse
under the self-gravity of baryonic gas and dark matter given by
Bertschinger (1985). It shows that an outgoing shock is always
formed during the infall of baryons.

A shock is actually a common feature of IGM fluid. Although the
cosmic baryonic gas is a Navier-Stokes fluid, the dynamic behavior
of the IGM is dominated by the gravity of the growth modes of the
dark matter. It has been recognized that the growth mode dynamics
of cosmic baryonic gas can be approximately described by the
random-force--driven Burgers equation (Gurbatov et al. 1989;
Vergassola et al. 1994; Berera \& Fang 1994; Jones 1999; Matarrese
\& Mohayaee 2002; Pando et al. 2004; He et al. 2004). The
dynamical behavior of a Burgers fluid depends on two
characteristic scales: (1) the correlation length of the random
force (in our case, the gravity of dark matter), and (2) the Jeans
length. When the former is larger than the latter, Burgers
turbulence develops in the baryonic gas in the non-linear regime.
That is, for the cosmic initial perturbations, which contain
fluctuations on small as well as larger scales, the Burgers
turbulence will definitely develop in nonlinear regime. The
Burgers turbulence is qualitatively different from Navier-Stokes'
turbulence. The latter generally consists of vortices on various
scales, while the former is a collection of shocks (Polyakov 1995;
Bouchaud et al. 1995; Yakhot 1998; L\"assig 2000; Davoudi et al.
2001). These features arise because the velocity field is
irrotational. Thus, the IGM velocity field in the nonlinear regime
can be understood as a field consisting of shocks.

Gas should slow down after passing through shocks, and postshock
gas should have higher density. On the other hand, dark matter is
not affected by the shocks. The nonuniform distribution of baryon
fraction is then caused by the postshock slowdown of gas.
Consequently, $F_b$ will be larger than 1 in the regions of
postshock, which can be in high-density ($\rho/\bar{\rho} >5$) and
low-density ($\rho/\bar{\rho} \simeq$ 1 - 5) regions.

Since the initial density perturbations of dark matter contain
components that have correlation scales larger than the Jeans
length at low density regions, the Burgers turbulence and the
shocks can happen in low as well as high overdense regions. This
property has also been shown in simulation results of He et al.
(2004), who found that the shock heating is significant in the
density regions of $\rho/\bar{\rho} \simeq 1 - 5$ at redshift
$z\simeq 0$. Therefore, the postshock slowdown mechanism will take
place in both high and low overdense regions.

Considering gas is moving from low- to high-density regions, the
postshock slowdown mechanism leads to the distributions of
$f_b({\bf x})$ having the following features.

1. Since all massive halos, at whatever redshift, formed by
gravitational collapse, the baryon shortage in high overdense
regions should be common at low as well as high redshifts.

2. Since baryons are detained on the way from low-density to
high-density (collapsed) regions, the $f_b({\bf x}) > f_{c}$
regions should be located in the low or moderate overdense areas
surrounding high overdense area, like massive halos.

3. Since baryons are detained in the postshock regions, the
$f_b({\bf x}) > f_{c}$ regions tend to be located in the higher
temperature regions.

4. Since baryons are frequently hampered by shocks, the velocities
of baryonic gas are statistically lower than those of dark matter.

The last point has been analyzed in Pando et al. (2004), who
showed that the probability distribution of baryons with large
peculiar velocities is much less than that of dark matter.

\section{Simulation samples}

The hydrodynamic equations of the baryonic gas in the universe is
the typical Navier-Stokes equation (Feng et al. 2004). As
discussed in the last section, the baryonic gas in the nonlinear
regime is characterized by (1) regions with discontinuities and
strong shocks and (2) regions with smooth and simple variations of
the field between the discontinuities. Therefore, an optimal
simulation scheme should be effective at capturing shock and
discontinuity transitions and in the meantime, at calculating
piecewise smooth functions with a high resolution.

For these reasons we do not use numerical schemes based on
smoothed particle hydrodynamic (SPH) algorithms. It is well known
that one of the main challenges to the SPH scheme is how to handle
shocks or discontinuities, because the nature of SPH is to smooth
the fields considered (e.g., B{\o}rve et al. 2001; Omang et al.
2003). Instead, we apply an Eulerian approach to simulate the IGM.
Among the popular algorithms of high-resolution shock capturing
are the total variation diminishing (TVD) scheme (Harten 1983) and
the piecewise parabolic method (PPM; Collella \& Woodward 1984).
Both schemes start from the integral form of conservation laws of
Euler equations and compute the flux vector based on cell averages
(finite volume scheme). These methods are able to produce
relatively sharp, nonoscillatory shock transitions. However, the
TVD scheme generally degenerates to first-order accuracy at
locations of smooth extrema (Godlewski \& Raviart 1996), and this
problem is serious in calculating the difference between
hydrodynamic quantities on both sides of the shock when the Mach
number of a gas is high. This is exactly the case that is
encountered in the gravitationally coupled IGM and dark matter
system.

Later, the essentially nonoscillatory (ENO) and then the weighted
essentially nonoscillatory (WENO) schemes were proposed as an
improvement over the TVD and PPM schemes (Harten et al. 1986; Shu
1998; Fedkiw et al. 2003; Shu 2003). It has been shown that for
solving problems governed by the high Reynolds number
Navier-Stokes equations, the WENO is more efficient than the TVD
and PPM schemes (Shi et al. 2003; Zhang et al. 2003). The WENO has
been successfully applied to problems of astrophysical
hydrodynamics, including stellar atmospheres (Del Zanna et al.
1998), high Reynolds number compressible flows with supernovae
(Zhang et al. 2003), and high Mach number astrophysical jets
(Carrillo et al. 2003). In the context of cosmological
applications, the WENO scheme has proved especially adept at
handling the Burgers equation (Shu 1999). Hence, we believe that
the WENO scheme would be effective at studying the problems
sensitive to strong shocks during the cosmological gravitational
clustering.

Recently, a hybrid hydrodynamic/$N$-body code based on the WENO
scheme was developed and has passed typical reliability tests, such
as the Sedov blast wave and the formation of Zel'dovich
pancakes (Feng et al. 2004). The code has been tested with
capturing gravitational shocks during the large-scale structure
formation (He et al. 2004). The WENO algorithm on cosmological
problems can be found in Feng et al. (2004).

For the purpose of this paper, we use the same simulation samples
as in He et al. (2004). The simulations were performed in a
periodic, cubic box of size 25 $h^{-1}$Mpc with a 192$^3$ grid and
an equal number of dark matter particles. We use the
clouds-in-cell method for mass assignment and interpolation and
adopt the seven-point finite difference to approximate the Laplacian
operator. The simulations start at redshift $z=49$, and the
results are output at redshifts $z$=6, 4, 3, 2, 1, 0.5, and 0. The
atomic processes, including ionization, radiative cooling, and
heating, are modeled similarly as in Cen (1992) in a plasma of
hydrogen and helium of primordial composition ($X=0.76$,
$Y=0.24$). Processes such as star formation and feedback due to
SNe and AGN activities are not taken into account yet. A uniform
UV background of ionizing photons is assumed to have a power-law
spectrum of the form $J(\nu) =J_{21}\times10^{-21}
(\nu/\nu_{HI})^{-\alpha}$ergs s$^{-1}$cm$^{-2}$sr$^{-1}$Hz$^{-1}$,
with parameters $J_{21}=1.0$ and $\alpha=1$. The photoionizing flux
is suddenly switched on at $z > 10$ to heat the gas and reionize
the universe.

For statistical studies, we randomly sampled 500 one-dimensional
fields from the simulation results at each redshift. Each
one-dimensional sample, of size $L$=25 $h^{-1}$ Mpc, contains 192
data points. For each of these points, information about the mass
density and peculiar velocity of dark matter, as well as the mass
density, peculiar velocity, and temperature of the IGM, is
recorded. We emphasize that, since we mostly focus on one-point
statistics, our results are dependent only on the fair sampling of
these data points, which is not relevant to the sample dimensions.

\section{Statistical properties of the baryon fraction}

\subsection{An example of the baryon fraction distribution}

Figure 1 is an example of the one-dimensional spatial distribution
of $F_b(x)$, which is the normalized baryon fraction
$F_b(x)=f_b(x)/f_{c}$, and $x$ is the comoving coordinate. We also
show in Figure 1 the mass densities $\rho_{DM}$ and $\rho_{IGM}$
of dark matter and baryonic matter and the temperature $T$ of the
baryons. The mass densities $\rho_{DM}$ and $\rho_{IGM}$ are in
units of their $\overline{\rho}_{DM}$ and $\overline{\rho}_{IGM}$,
respectively. Figure 1 clearly shows that $F_b(x)$ is
significantly nonuniform. Some small-scale deviation of $F_b(x)$
from unity can be explained by the Jeans smoothing. However, there
are deviations on scales of one to a few $h^{-1}$ Mpc, which is
larger than the corresponding Jeans length of the baryonic gas.

In Figure 1 the mass density on the left side of the simulation
box is higher than on the right side, and from the comparison of
density distributions at $z=0.5$ and $0$, one can see that matter
is infalling from the right to the left region. We see a shock at
about $x=11$ $h^{-1}$ Mpc at $z=0.5$. The temperature of preshock
gas is $\sim$10$^{3}$ K and increases to $\sim$10$^{7}$K after
passing through the shock. That is, the temperature increases by a
factor of $\sim 10^4$. According to the shock theory of a
polytropic gas with index $\gamma$ (Landau \& Lifshitz 1959), the
Mach number should be $M \simeq \sqrt{10^4}\simeq 100$. This value
is reasonable, since under such a strong shock, the density of
baryons is enhanced by a factor of $\sim (\gamma+1)/(\gamma-1)=
4$, while the mass density of dark matter is not affected by the
shock. Therefore, $F_b(x)$ can be as high as 4. The positions of
this shock at time $z=0.5$ and $0$ indicate that it is an outgoing
shock with respect to the high-density area. At $z=0.5$, the peak
of $F_b(x)$ is just located on the postshock side or
high-temperature side. At $z=0$, more high peaks of $F_b(x)$
develop behind the shock. Namely, more baryonic matter is detained
by the shock. This result is consistent with the picture of
postshock slow down of the baryon flow described in \S 2.

This picture can be more clearly seen in Figures 2 and 3. Figure 2
presents two-dimensional contours of the baryonic gas and dark
matter densities. One can see at the lower part of the plots that
a massive halo on the scale of $\sim$5 $h^{-1}$ Mpc is formed. The
$\rho_{IGM}>1$ region is obviously larger than the dark matter
counterpart at $\rho_{DM}>1$. This means that more baryons remain
in the low overdense area. This discrepancy cannot be caused by
the Jeans diffusion. Figure 3 gives the two-dimensional contours
of temperature and $\rho_{IGM}/\rho_{DM}$ (i.e., $F_b$) of the
same slice in Figure 2. Figure 3 shows that the size of the
high-temperature region is about the same as that of the
$\rho_{IGM}>1$ region. From the right panel of Figure 3, we see
that the contours of $F_b>1$ are located outside the center of the
object. Conversely, the central part of this object, where the
mass density is higher than that at the boundary, has only
$F_b<1$. This feature is very common. All the clustered structures
like the one in Figures 2 and 3 have $F_{b}<1$ in their center and
are surrounded by $F_b>1$ regions. That is, baryons accumulate in
low and moderate (dark matter) overdense areas.

\subsection{Baryon fraction - density and baryon fraction - temperature
relations}

In Figure 4, we show the scatter plots between the baryon fraction
and the dark matter density at redshifts $z$=6, 4, 2, 1, 0.5, and
0. Each panel in Figure 4 contains 19,200 data points of the
randomly drawn 500 one-dimensional samples at each redshift. It is
expected that, in each panel, most of the data points are
distributed around $F_b\simeq 1$. If baryons underwent only linear
evolution, all the points should be on the line of $F_b\simeq 1$
for all $\rho_{DM}$. However, the scatters around $F_b = 1$ are
significant, among which the relatively small scatters can be
explained by the Jeans smoothing, while the points with $F_b \gg
1$ should be attributed to strong shocks or discontinuity
transitions. These points are especially prominent in the regions
of $\rho_{DM} <5$.

We can see from Figure 4 that the scatter of $F_b$ at redshifts
$z\simeq 2$ is as significant as that at low redshifts. Only at
$z\geq 4$ does the $F_b$ scatter become small. This is consistent
with the prediction in \S 2 that the deviation of $F_b$ from 1 at
early times should be as prominent as that at later times. This
feature is much different from the virialization of baryons in
collapsed massive halos. The virialization process generally takes
place during the formation of the halos, while the shortage of
baryons in massive halos proceeds prior to the formation of the
halos.

Figure 5 gives the relation between the baryon fraction and the
dark matter density at $z=0$ for baryon and dark matter fields
smoothed on scales 0.26, 1.04, and 4.17 $h^{-1}$ Mpc, respectively
(the method of smoothing is given in Appendix A). By this
treatment, the linear size of each cell will increase by factors
of $2$, $2^{3}$, and $2^{5}$, respectively. We see that the
scatter of the smoothed $F_b$ on the scale 1.04 $h^{-1}$ Mpc
remains nearly the same as that on the scale 0.13 $h^{-1}$ Mpc
even when $\rho_{DM}< 1$. Since the size of high-$F_b$ regions is
generally larger than 1 $h^{-1}$ Mpc (see Figs 1, 2, and 3), the
scatter cannot be erased by smoothing on scales of about 1
$h^{-1}$ Mpc. Therefore, the scatter is intrinsic.

As a quantitative comparison for the dependence of $F_b$ on the
dark matter density, in Table 1 we list the mean baryon fraction
in each density interval at several redshifts. We see from Table 1
that at all redshifts from 0 to 4, the mean $F_b$ is always in the
range of 0.7 to 0.9 for the very high densities. Since such
high-density areas are the possible sites for the formation of
galaxy clusters, this result provides a cosmological explanation
of the shortage of baryons in dense dark halos, like galaxy
clusters. We can see that the baryon fraction of clusters can be
as low as $\sim0.8$ in general.

In Figure 6, we show the relations between baryon fraction and
temperature of baryonic gas at redshifts $z$=4, 2, 1, and 0. Each
panel in Figure 6 contains 19,200 data points of the randomly
drawn 500 samples at each redshift. The $z$-evolution of $F_b$ in
$T$-space is different from that of $\rho_{DM}$-space. In Figure 4
the high-$F_b$ points are always in the density range $0.5 <
\rho_{DM} <5$ for all redshifts considered, while in Figure 6, the
high-$F_b$ points are mostly located at $10^4 <T < 10^5 K$ for
$z=4$ and $T>10^5$ K for $z=0$. That is, the baryonic gas with
$F_b>1$ generally tends to lie in high-temperature ($T>10^5$ K)
areas. This is consistent with the picture that strong shocks can
reduce the flow of the baryons in postshock areas and meanwhile
increase the gas temperature by a factor of $10^2$-$10^4$ (Fig.
1).

\subsection{High baryon fraction phase (HBFP)}

From Figures 4 and 6, we see that almost all the baryonic gas with
$F_b > 1$ is hot ($T>10^5$ K) and moderately dense ($\rho_{DM} <
5$). From these results, we can define a special phase of baryonic
gas with the indicator $F_b > 1$: it can be called high baryon
fraction phase (HBFP). Baryonic gas in the HBFP is characterized
by two properties: (1) it is located mostly in the moderately
dense regions with $\rho_{DM} < 5$, and (2) its temperature is
larger than $10^5$ K (see \S 4.2). Conversely, Figure 7 indicates
that almost all the gas with $\rho_{DM} < 5$ and $T>10^5$ K has
mean $F_b$ larger than 1. Figure 8 is another version of Figure 7,
which shows the $\rho_{DM}$-dependence of the mean baryon fraction
for baryonic gas with $T <10^5$ K and $T>10^5$ K. It also
indicates that almost all the gas with $T>10^5$ K and $\rho_{DM} <
5$ are $F_b > 1$. More precisely, 89\% baryons in the $T>10^5$ K
and $\rho_{DM} < 5$ regions are $F_b>1$. Thus, the phase of
$F_b>1$ is approximately equivalent to the thermodynamic condition
$T>10^5$ K and $\rho_{DM} < 5$. The HBFP is a special phase of
baryonic gas formed as a result of the nonlinear evolution of
baryon--dark matter systems.

Figure 9 is the same as the $z=0$ panel of Figure 8, but with data
smoothed on scales 0.26, 1.04, and 4.17 $h^{-1}$ Mpc. All the
basic features of Figure 9 are the same as those of Figure 8. That
is, $\overline{F_b}<1$ is generally in regions with high density
and high temperature, while $\overline{F_b}>1$ is mostly in
$T>10^5$ K, but with lower $\rho_{DM}$. In the range $0.03
<\rho_{DM}<1$, the curves of $F_b$ in Figure 9 are clearly less
sensitive to smoothing scales. This shows that shot noise does not
affect our conclusions about the mean baryon fractions (see
Appendix B).

On large enough scales, the cosmic clustering can still remain in
the linear regime, and therefore one may expect that the
inhomogeneity of baryon fraction distribution would disappear on
large scales. The mean baryon fraction within a large radial
region around a cluster has to be asymptotically approaching 1.
Figure 9 shows that $\overline{F_b}$ is still quite inhomogeneous
even at a scale as large as 4.17 $h^{-1}$ Mpc. Therefore, the
asymptotic radius for $F_b \rightarrow 1$ should be larger than 4
$h^{-1}$ Mpc.

In terms of observations, the HBFP is very different from other
phases of the baryonic gas. As for the mass density, the HBFP is
about the same as the IGM for Ly$\alpha$ forests, which is given
by the absorption of HI in the regions $0.5 < \rho_{DM} < 5$ (Bi
\& Davidsen 1997). However, the HBFP cannot be detected by the
Ly$\alpha$ forests of QSO's absorption spectrum, because the
fraction of $HI$ is too low to be seen when temperature $T$ is
higher than $10^5$ K. As for the temperature, the HBFP is about
the same as the so-called WHIM (warm hot IGM) (Cen \& Ostriker
1999), which is generally defined as baryonic gas with temperature
$10^5-10^7$ K, located in the regions close to the filaments, with
mass density $5 <\rho_{DM}<200$ (Dav\'e et al. 2001). The WHIM
would be a source of soft X-ray background. However, the number
density of the HBFP gas is too low for it to be a source of soft
X-ray emissions, and hence it cannot be detected via the soft
X-ray observations either. For these reasons, the HBFP baryons are
out of the baryon budget counted with the current observations.
The contribution of the HBFP to the total cosmic baryon budget is
given in Figure 10. The mass fraction of HBFP in the baryon budget
is about 2.5\% at $z=$ 3 and increases to 14.4\% at the present
day ($z=0$). Therefore, the HBFP would not be a negligible
component of the missing baryons.

\section{Conclusions and Discussions}

In the nonlinear regime, a system consisting of collisional
baryons and collisionless dark matter is generally characterized
by strong shocks and discontinuities in the baryonic fluid. The
shocks outgoing from high-density regions generally slow down the
infall motion of baryons from low-density to high-density regions.
Consequently, the baryon fraction in lower overdense areas is
higher than the cosmic value, while in higher overdense areas it
is lower than the cosmic value. We use $N$-body/hydrodynamic
simulation samples produced by the WENO code to quantitate the
deviation of the baryon fraction from cosmic value. We conclude
that the overall baryon matter of massive halos like clusters is
lacking by about 10\%--20\%. We also find that the HBFP is
composed of the gas with $T>10^5$K and is located in
moderate-density ($\rho_{DM}<5$) regions.

Thus, in terms of the baryon fraction, gas in massive halos, like
clusters, is in the phase of low baryon fraction, while gas with
$\rho_{DM}<5$ and $T>10^5$K is in HBFP. Both the low baryon
fraction phase and HBFP are formed by the same mechanism of the
shock-caused separation between the baryonic gas and dark matter.
About 14\% baryons in the universe today are presumably hidden in
the HBFP. The HBFP can be traced neither by QSO absorption
spectrum nor by X-ray emissions. However, the ionized electrons in
the HBFP would be capable of scattering CMB photons and might
generate secondary cosmic temperature fluctuations. The
Sunyaev-Zel'dovich effects might be promising for detecting the
existence of the HBFP.

We should point out that star formation and their feedback on the
IGM evolution are not considered in our simulation. Generally
speaking, there are two types of feedbacks: (1) photoionization
heating by the UV emission of stars and AGNs and (2) injection of
hot gas and energy by supernova explosions or other sources of
cosmic rays. The photoionization heating actually can be properly
considered, if the UV background is adjusted by fitting the
simulation with the observed mean flux decrement of QSO Ly$\alpha$
absorption spectra (Feng et al. 2003). The effect of injecting hot
gas and energy by supernovae is localized in massive halos, and
thus it may change some results with clusters but does not affect
the IGM in low- and moderate-density areas. Therefore, the
features of the HBFP would not be significantly affected even if
considering the effect of star formation.

\acknowledgments

The authors thank David Weinberg for his very helpful suggestions
and comments in his referee's report. P.H. is supported by a
Fellowship of the World Laboratory. L.L.F. acknowledges support
from the National Science Foundation of China (NSFC) and National
Key Basic Research Science Foundation. This work is partially
supported by the National Natural Science Foundation of China
(10025313) and the National Key Basic Research Science Foundation
of China (NKBRSF G19990752).

\appendix

\section{Smoothing with scaling functions}

Consider a one-dimensional density fluctuation $\delta(x)$ on a
spatial range from $x=0$ to $L$. We divide the space into $2^j$
segments labeled $l=0,1,...2^j-1$, each of size $L/2^j$. The index
$j$ is a positive integer and gives length scale $L/2^j$. The
larger the value of $j$, the smaller the length scale. The index
$l$ represents the position and corresponds to the spatial range
$lL/2^j < x < (l+1)L/2^j$. Hence, the space $L$ is decomposed into
cells $(j,l)$.

The discrete wavelet is constructed such that each cell $(j,l)$
supports a compact function, the scaling function $\phi_{j,l}(x)$.
In our calculations, the Daubechies 4 (D4) wavelet (Daubechies
1992) is used. The scaling function satisfies the orthogonal
relation
\begin{equation}
\int \phi_{j,l}(x)\phi_{j,l'}(x)dx = \delta^K_{l,l'},
\end{equation}
where $\delta^K$ is Kronecker delta function. The scaling function
$\phi_{j,l}(x)$ is a window function on scale $j$ centered at the
segment $l$. The normalization of the scaling function is $\int
\phi_{j,l}(x)dx=(L/2^j)^{1/2}$.

For a field $\rho(x)$, its mean in cell $(j,l)$ can be estimated by
\begin{equation}
\rho_{j,l}=\frac{\int_{0}^{L} \rho(x)\phi_{j,l}(x)dx}
   {\int_{0}^{L} \phi_{j,l}(x)dx}=
   \left (\frac{2^j}{L}\right)^{1/2} \epsilon^{\rho}_{j,l},
\end{equation}
where $\epsilon^{\rho}_{j,l}$ is called the scaling function
coefficient (SFC), given by
\begin{equation}
\epsilon^{\rho}_{j,l}= \int_{0}^{L} \rho(x)\phi_{j,l}(x)dx.
\end{equation}

A one-dimensional field $\rho(x)$ can be decomposed into
\begin{equation}
\rho(x) =
  \sum_{l=0}^{2^j-1}\epsilon^{\rho}_{j,l}\phi_{j,l}(x) + O(\geq j).
\end{equation}
The term $O(\geq j)$ in equation (A4) contains only the
fluctuations of the field $F(x)$ on scales equal to and less than
$L/2^j$. This term does not have any contribution to the window
sampling on scale $j$. Thus, for a given $j$, the one-point
variables $\rho_{j,l}$ or $\epsilon^{\rho}_{j,l}$ ($l=0,
1...2^j-1$) give a complete description of the field $F(x)$
smoothed on scale $L/2^j$. As one-point variables the
$\epsilon^{\rho}_{j,l}$ are similar to the measure given by
count-in-cell technique. However, the orthonormality equation (A1)
ensures that the set of $\rho_{j,l}$ or $\epsilon^{\rho}_{j,l}$
does not cause false correlations. When the ``fair sample
hypothesis'' (Peebles 1980) holds, the average over the ensemble
of the random field can be estimated by averaging over modes
$(j,l)$.

\section{Errors of Poisson sampling}

Consider a random field $\rho^M(x)=\bar{\rho}[1+\delta(x)]$, where
$\delta(x)=[\rho(x)-\bar{\rho}]/\bar{\rho}$, and $\bar{\rho}$ is
the mean. Obviously, $\langle\delta(x)\rangle =0$. The observed or
simulated field $\rho(x)$ is considered to be a Poisson sampling
of the field $\rho^M(x)$. The characteristic function of the
$\rho(x)$ is
\begin{equation}
Z[e^{i\int\rho(x)u(x)dx}]= \exp\left \{ \int dx
\rho^M(x)[e^{iu(x)}-1] \right \},
\end{equation}
and the statistic of $\rho(x)$ is given by
\begin{equation}
\langle\rho(x_1)...\rho(x_n)\rangle_P =\frac{1}{i^n} \left [ \frac
{\delta^n Z}{\delta u(x_1)... \delta u(x_n)} \right ]_{u=0},
\end{equation}
where $\langle ...\rangle_P$ is the average for the Poisson
sampling. We then have
\begin{equation}
\langle\rho(x)\rangle_P = \rho^M(x),
\end{equation}
and
\begin{equation}
\langle\rho(x)\rho(x')\rangle_P = \rho^M(x)\rho^M(x') +
\delta^D(x-x')\rho^M(x).
\end{equation}
Subjecting equations (B3) and (B4) to the transform equation (A2),
we have
\begin{equation}
\langle\rho_{j,l}\rangle_P= \rho^M_{j,l}
\end{equation}
and
\begin{equation}
\langle\rho_{j,l}\rho_{j,l}\rangle_P = \rho^M_{j,l}\rho^M_{j,l} +
\frac{2^j}{L}\int\rho^M(x)\phi_{j,l}(x)\phi_{j,l}(x)dx.
\end{equation}
Therefore, the Poisson sampling error of the measurement
$\rho_{j,l}$ can be estimated by
\begin{equation}
\sigma^2_{j,l}\equiv \langle\rho_{j,l}\rho_{j,l}\rangle_P-
   \langle\rho_{j,l}\rangle^2_P=
\frac{2^j}{L}\int\rho^M(x)\phi_{j,l}(x)\phi_{j,l}(x)dx.
\end{equation}
If we use the normalized density variable, equation (B7) gives
\begin{equation}
\sigma^2_{j,l}=\left \langle \frac{\rho_{j,l}}{\bar{\rho}}\
        \frac{\rho_{j,l}}{\bar{\rho}} \right \rangle_P-
\left \langle\frac{\rho_{j,l}}{\bar{\rho}} \right \rangle^2_P=
\frac{2^j}{L} \frac{1}{\bar{\rho}}+
\frac{2^j}{L}\int\frac{\delta(x)}{\bar{\rho}}\phi_{j,l}(x)\phi_{j,l}(x)dx.
\end{equation}
Since the Poisson samplings for different modes $(j,l)$ are
uncorrelated, the Poisson sampling error of the measurement of
$(1/N)\sum_{N}\rho_{j,l}$, which is the mean of $\rho_{j,l}$ over
$N$ modes, is
\begin{equation}
\frac{1}{N}(\sum_{N}\sigma^2_{j,l})^{1/2}= \frac{1}{N}\left
[N\frac{2^j}{L}\frac{1}{\bar{\rho}}+\sum_{N}
\frac{2^j}{L}\int\frac{\delta(x)}{\bar{\rho}}\phi_{j,l}(x)\phi_{j,l}(x)dx
\right ]^{1/2}.
\end{equation}
Considering that $\langle \delta(x)\rangle=0$, the second term on
the right-hand side generally is negligible, and we then have
\begin{equation}
\frac{1}{N}(\sum_{N}\sigma^2_{j,l})^{1/2}\simeq
\frac{1}{\sqrt{N}}\left(\frac{1}{(L/2^j)\bar{\rho}}\right )^{1/2},
\end{equation}
where $N$ is the number of modes. The factor $(L/2^j)\bar{\rho}$
is the mean mass, or mean number of particles in the cell $L/2^j$.
A smoothed field takes a larger $(L/2^j)$. Thus, if a statistical
result is weakly dependent on the $(L/2^j)$-smoothing, the effect
of Poisson error is negligible.

In some calculations, we only choose the modes for which the
density is restricted to a given range. The average over these
modes may not give $\langle \delta(x)\rangle=0$. In this case, the
$\delta(x)$ term in equation (B8) is not negligible. For instance,
if $\langle\delta(x)\rangle\simeq 100$, the error would be
increased by a factor of 10. Nevertheless, the contribution of the
$\delta(x)$ term is also proportional to $2^j/L$. If a statistical
average over $500$ modes with mean number density is larger than
$0.03$, the Poisson error will not be larger than 25\%. Therefore,
in Figures 7--9, the shot noise in the range of $\rho_{DM}
>0.03$ are negligible.

\newpage

\begin{figure*}[hbt]
\centerline{\psfig{figure=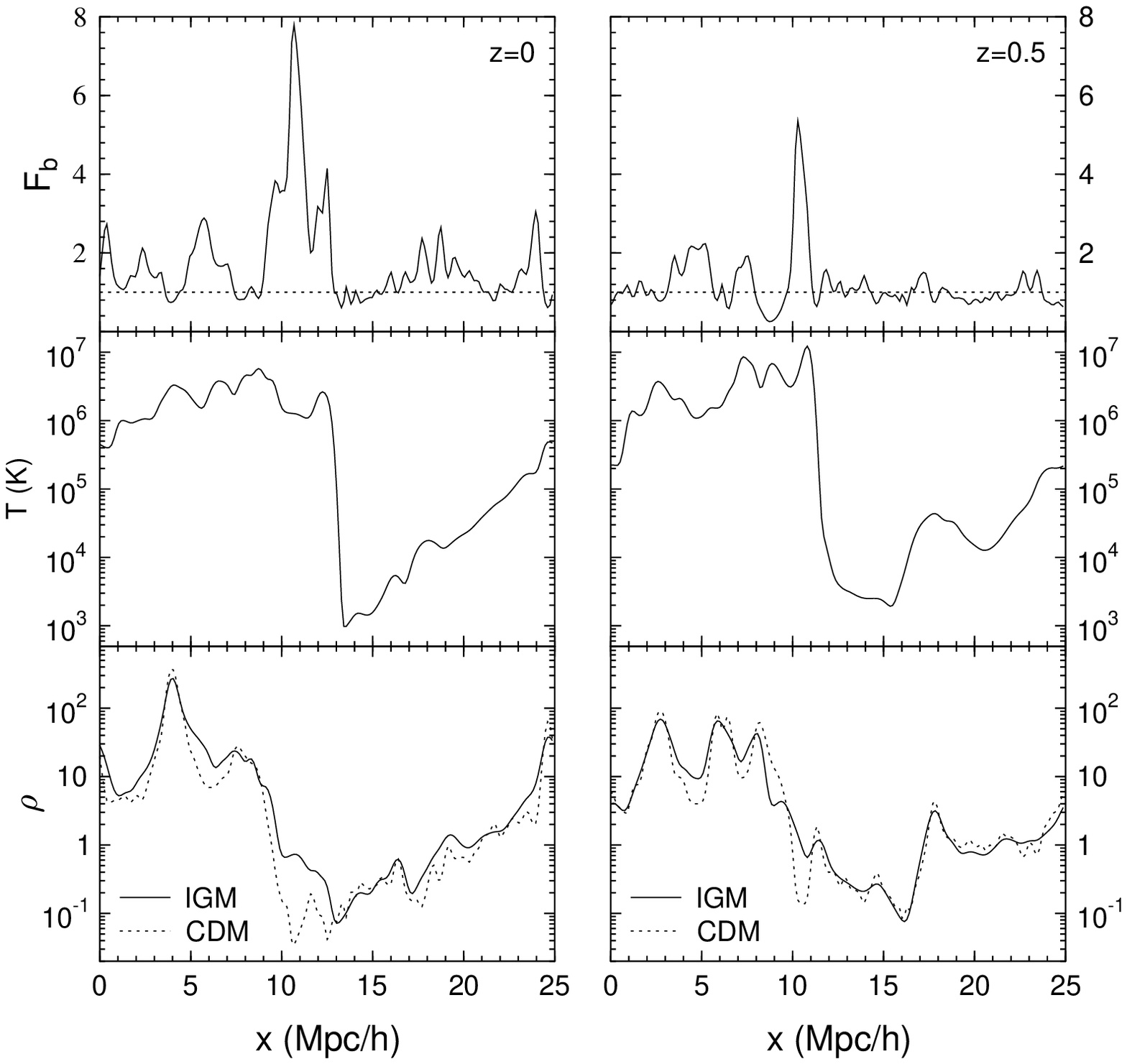}}

\caption{Example of the one-dimensional distribution of
         $F_b(x)$ at $z$=0 and 0.5, where $x$ is the comoving
         coordinate. The corresponding temperature and density
         fields are also shown in the bottom panels.}
\end{figure*}

\newpage

\begin{figure*}[hbt]
\centerline{\psfig{figure=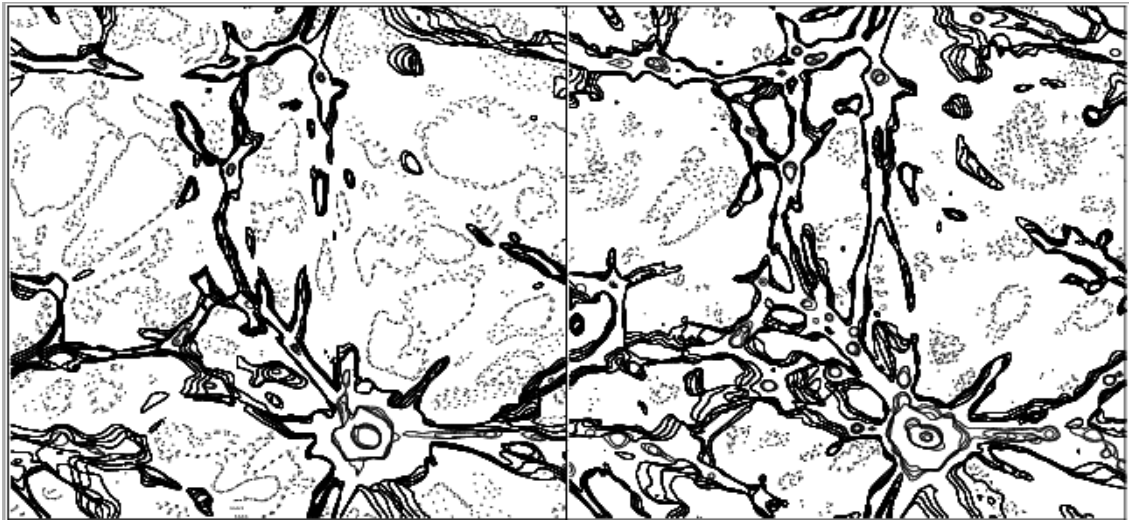,width=15cm}}

\caption{Density contour plots of dark matter (right panel) and
         baryonic gas (left panel) for a slice of 0.26 $h^{-1}$
         Mpc thickness at $z=0$. The solid contours encompass
         the overdense regions with $\rho=e^{i/2}$, $i=0,1,2...$,
         ($\bar{\rho}$ is normalized to 1). The dotted lines
         represent the boundaries of the underdense regions with
         $\rho=e^{-i/2}$, $i=1,2...$.}
\end{figure*}

\begin{figure*}[hbt]
\centerline{\psfig{figure=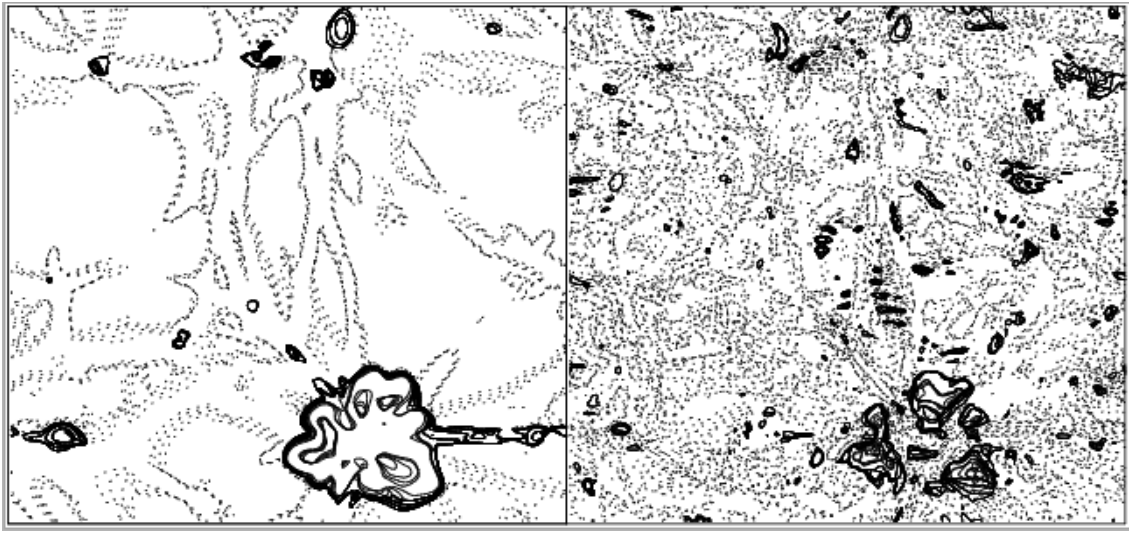,width=15cm}}

\caption{Contour plots of baryon-to-dark ratio $F_b$ (right
         panel) and temperature $T$ (left panel) for the same
         slice as in Fig. 2. The solid contours represent,
         respectively, the regions with $F_b = e^{i/3},
         i=0,1,2,...$ and $T=e^{i/2}\times 10^5$ K, $i=0,1,2,...$.
         The dotted lines represent $F_b = e^{-i/3},
         i=1,2,...$ and $T=e^{-i/2}\times 10^5$ K, $i=1,2,...$
         regions.}
\end{figure*}

\newpage

\begin{figure*}[hbt]
\centerline{\psfig{figure=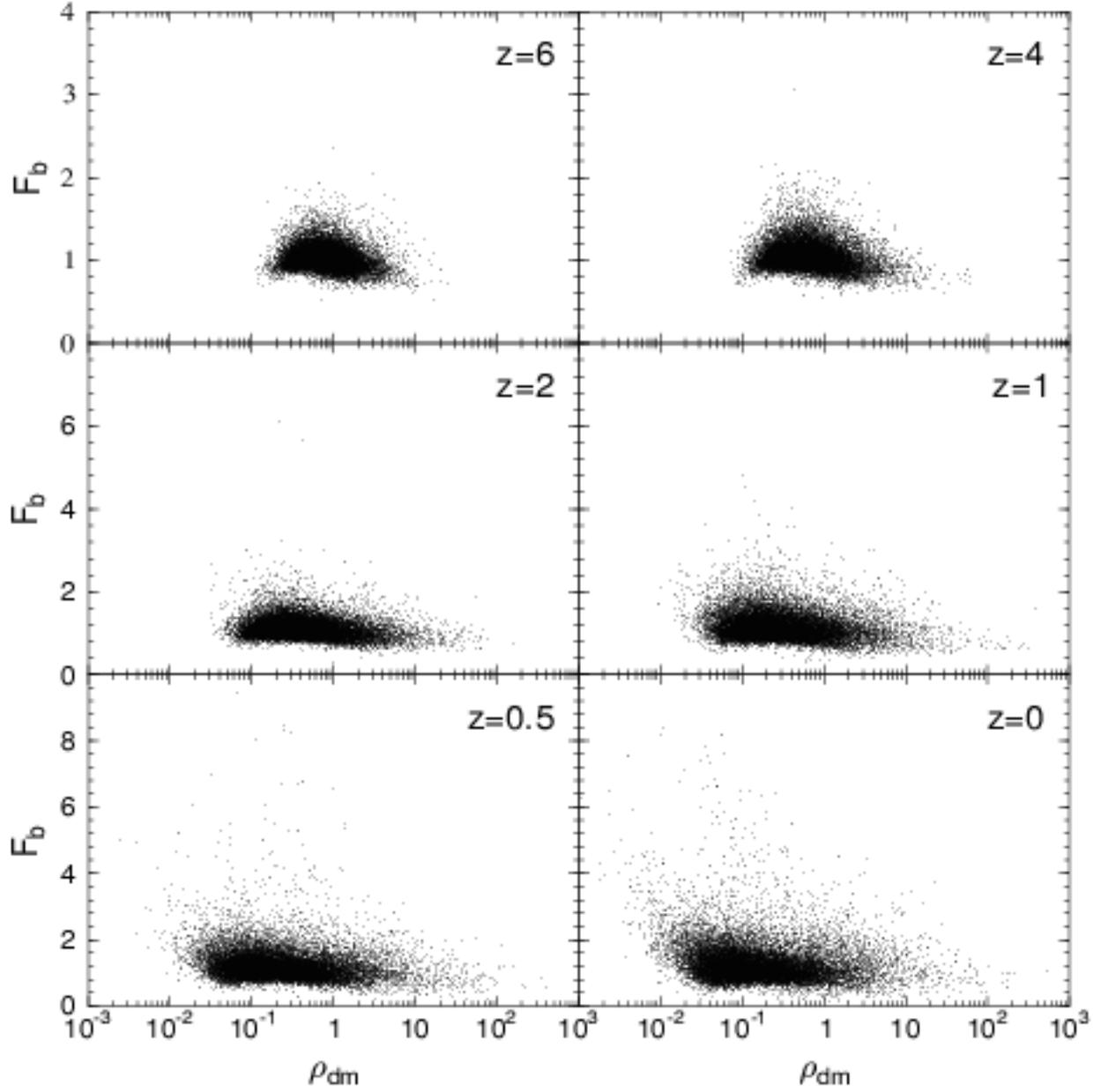}}

\caption{Relation between $F_b(x)$ and $\rho_{DM}(x)$
         for each point $x$ at redshifts $z$=6, 4, 2, 1, 0.5,
         and 0. In each panel, the data consist of 19,200
         points randomly drawn from the simulation samples.
         The density $\rho_{DM}$ is in units of ${\bar\rho}_{DM}$.}
\end{figure*}

\newpage

\begin{figure*}[hbt]
\centerline{\psfig{figure=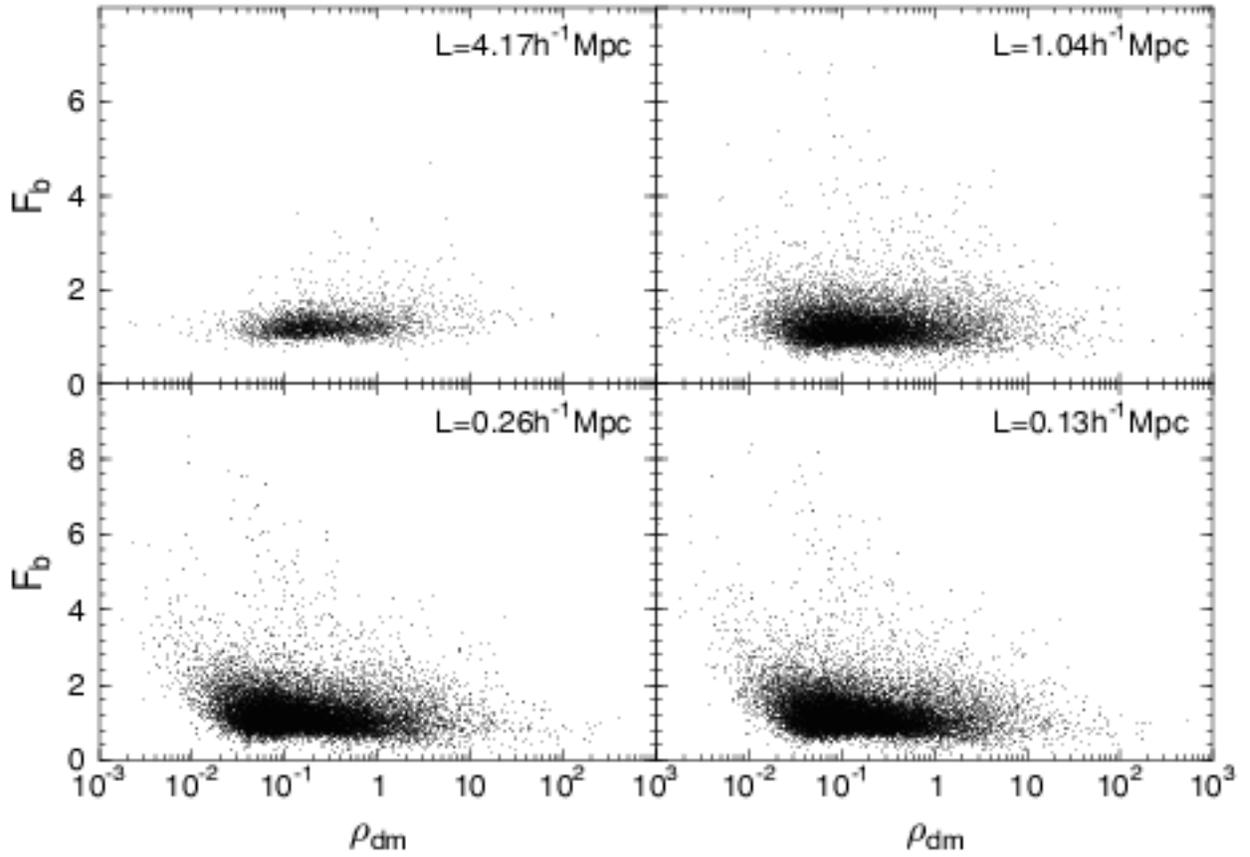}}

\caption{Relation between $F_b$ and $\rho_{DM}$
         for data smoothed on scales 0.13, 0.26, 1.04,
         and 4.17 $h^{-1}$ Mpc at $z=0$.}
\end{figure*}

\newpage

\begin{figure*}[hbt]
\centerline{\psfig{figure=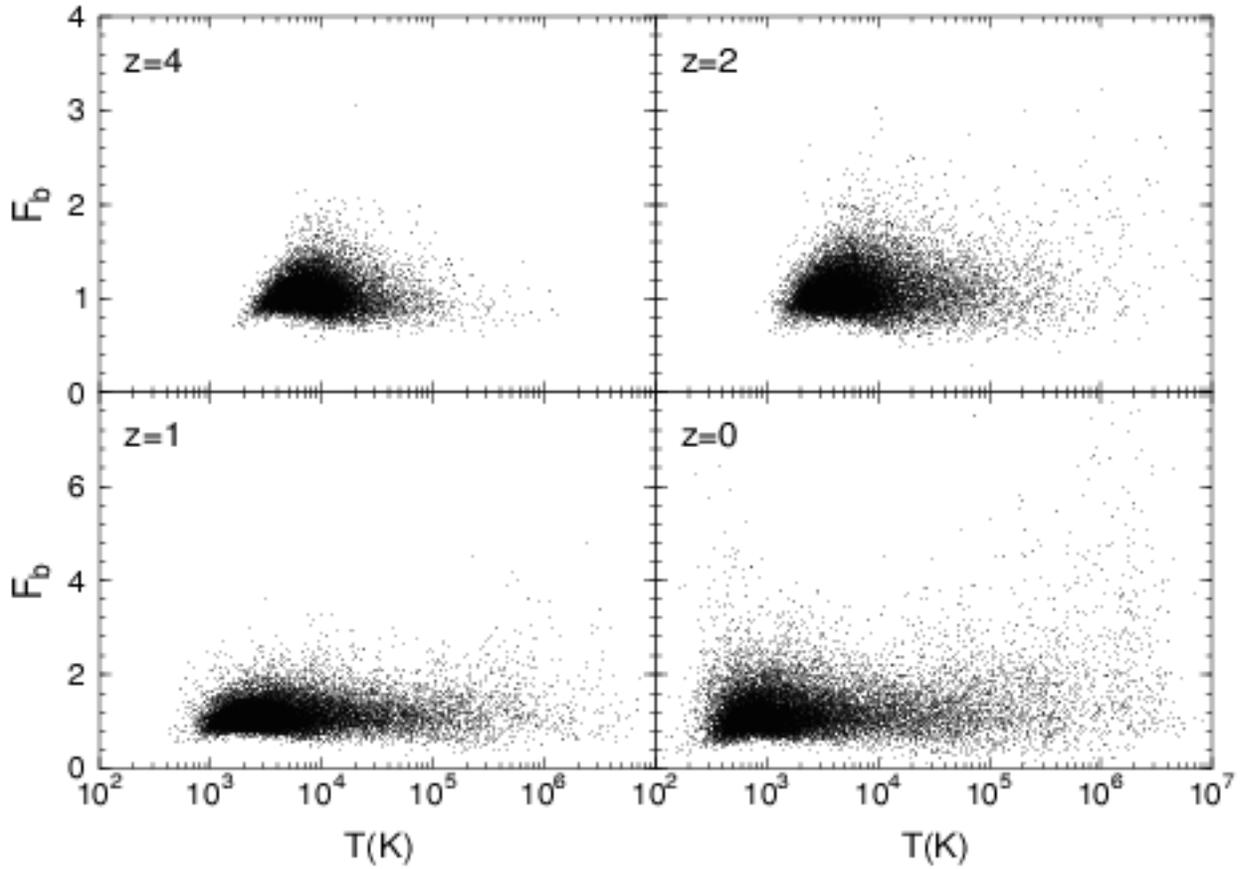}}

\caption{Relation between $F_b(x)$ and temperature $T(x)$
         for each point $x$ at redshifts $z$=4, 2, 1, and 0.
         The data consist of 19,200 points randomly drawn
         from the simulation samples.}
\end{figure*}

\newpage

\begin{figure*}[hbt]
\centerline{\psfig{figure=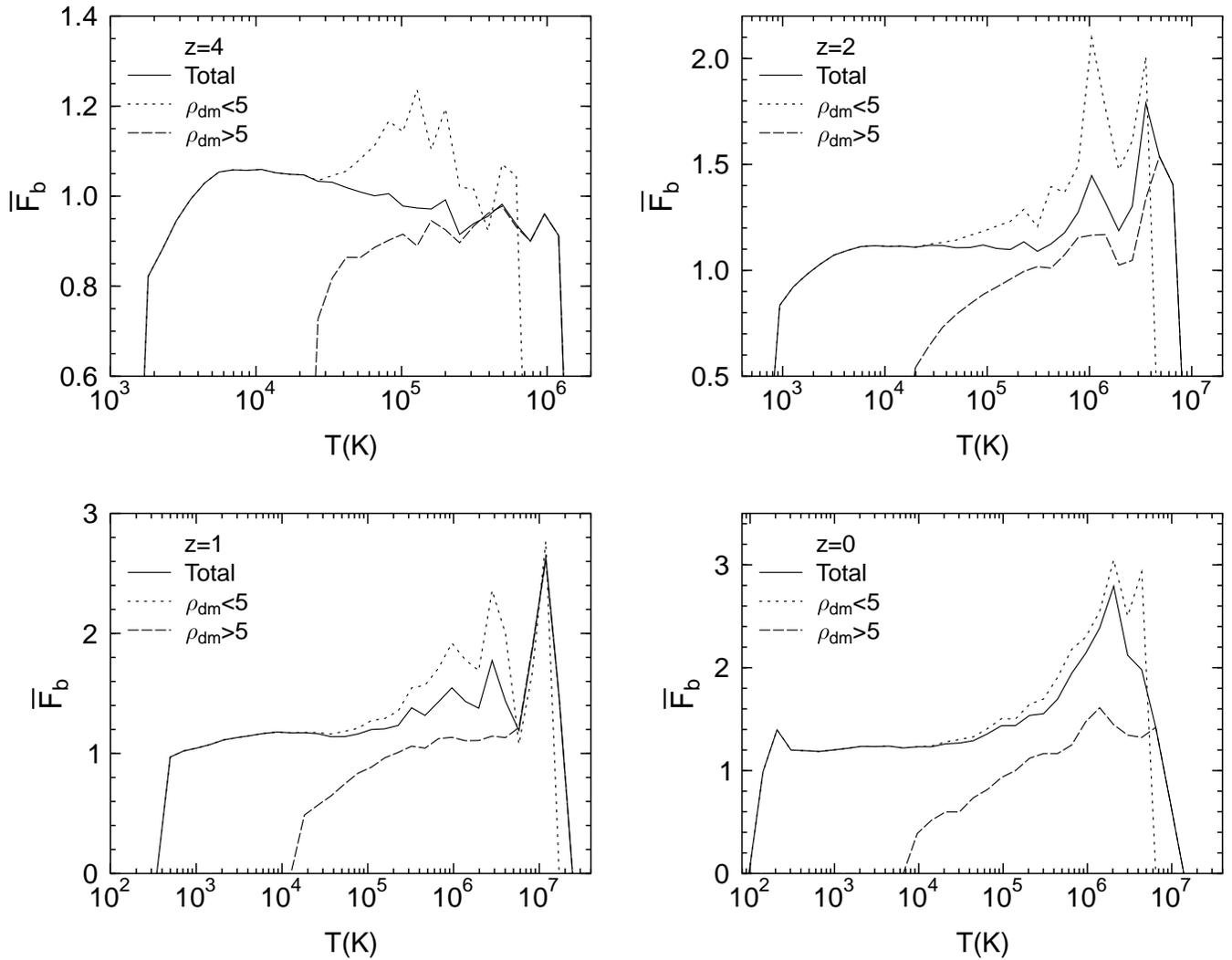}}

\caption{Averaged $F_b$ vs. temperature $T$. The three
         curves are for (1) the total samples (solid lines),
         (2) the data with $\rho_{DM}<5$ (dotted lines),
         and (3) the data with $\rho_{DM}>5$ (dashed lines).}
\end{figure*}

\newpage

\begin{figure*}[hbt]
\centerline{\psfig{figure=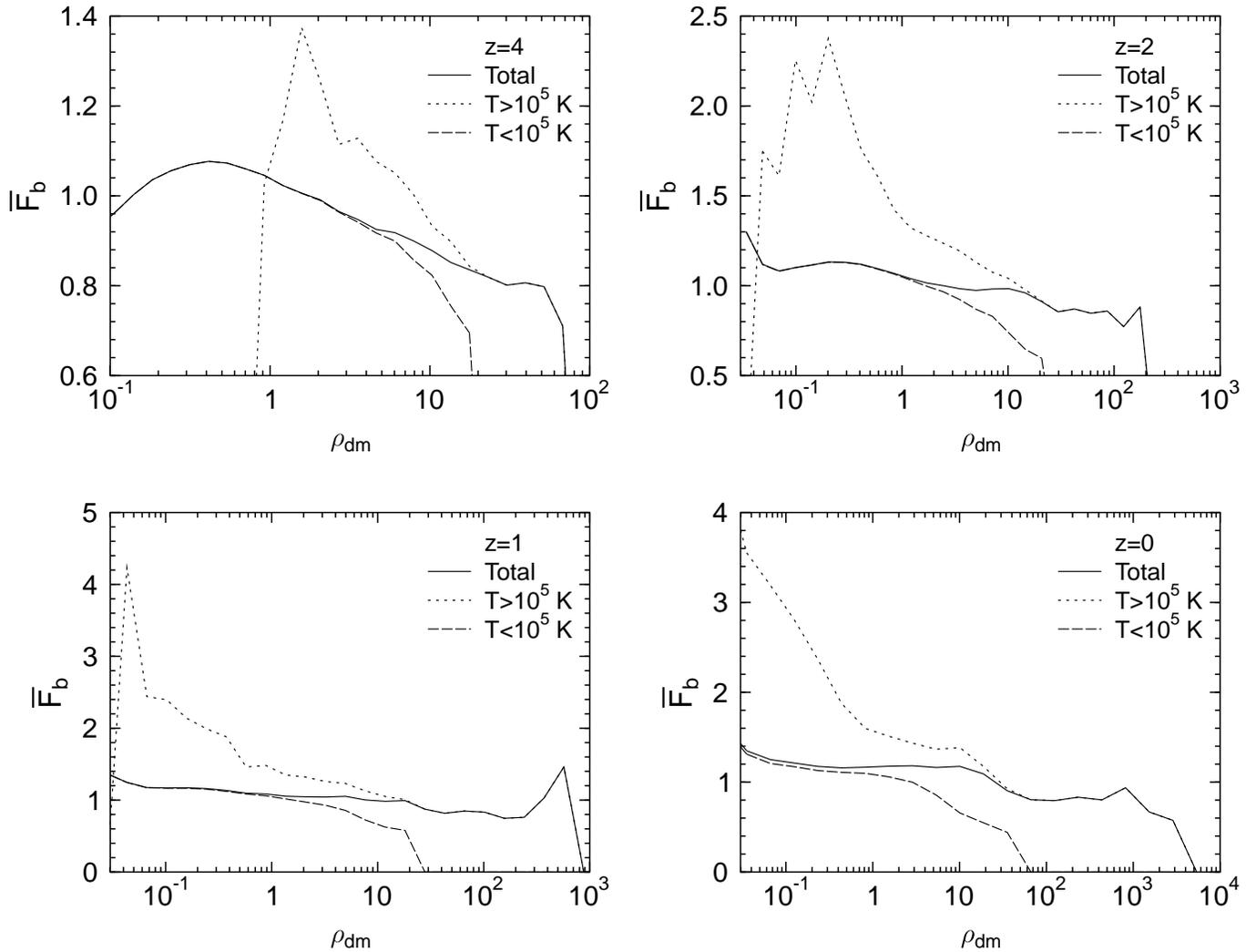}}

\caption{Averaged $F_b$ vs. $\rho_{DM}$. The three curves
         are for (1) the total samples (solid lines), (2) the
         data with temperature $T<10^5$ K (dashed lines), and
         (3) the data with $T>10^5$ K (dotted lines).}
\end{figure*}

\newpage

\begin{figure*}[hbt]
\centerline{\psfig{figure=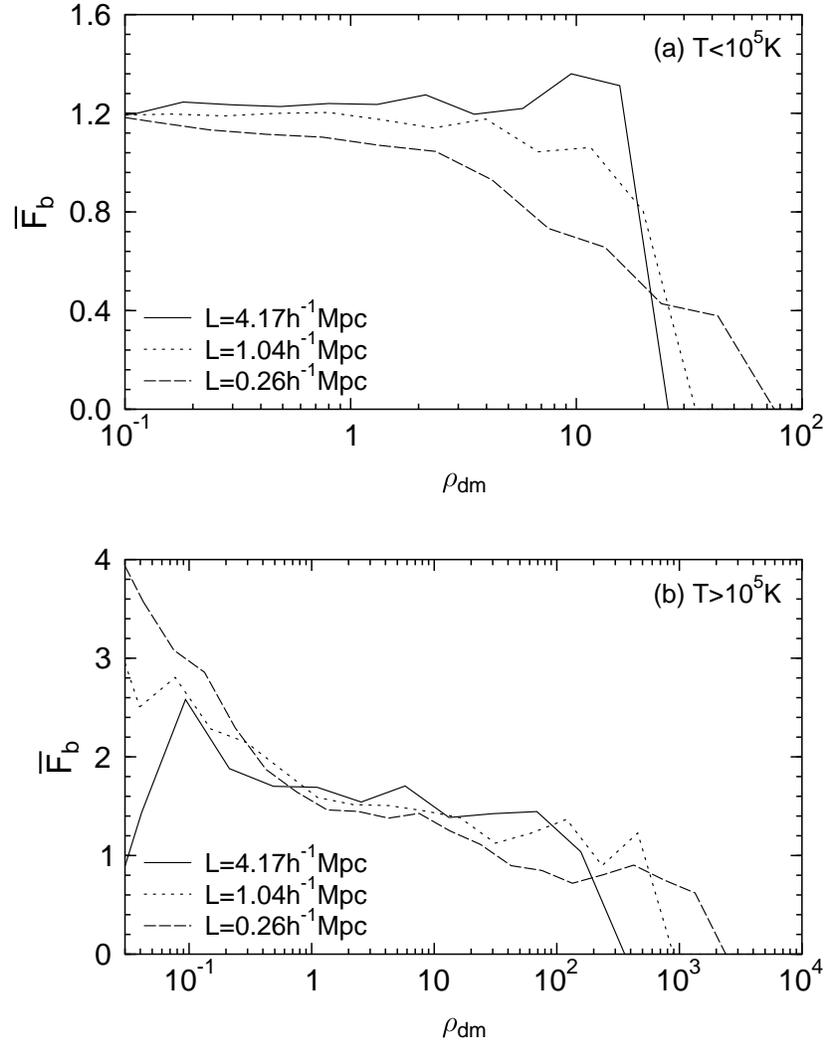}}

\caption{Same as Fig. 8, but only for $z=0$, and the data
         are smoothed on scales of 0.26, 1.04, and 4.17 $h^{-1}$
         Mpc, respectively.}
\end{figure*}

\newpage

\begin{figure*}[hbt]
\centerline{\psfig{figure=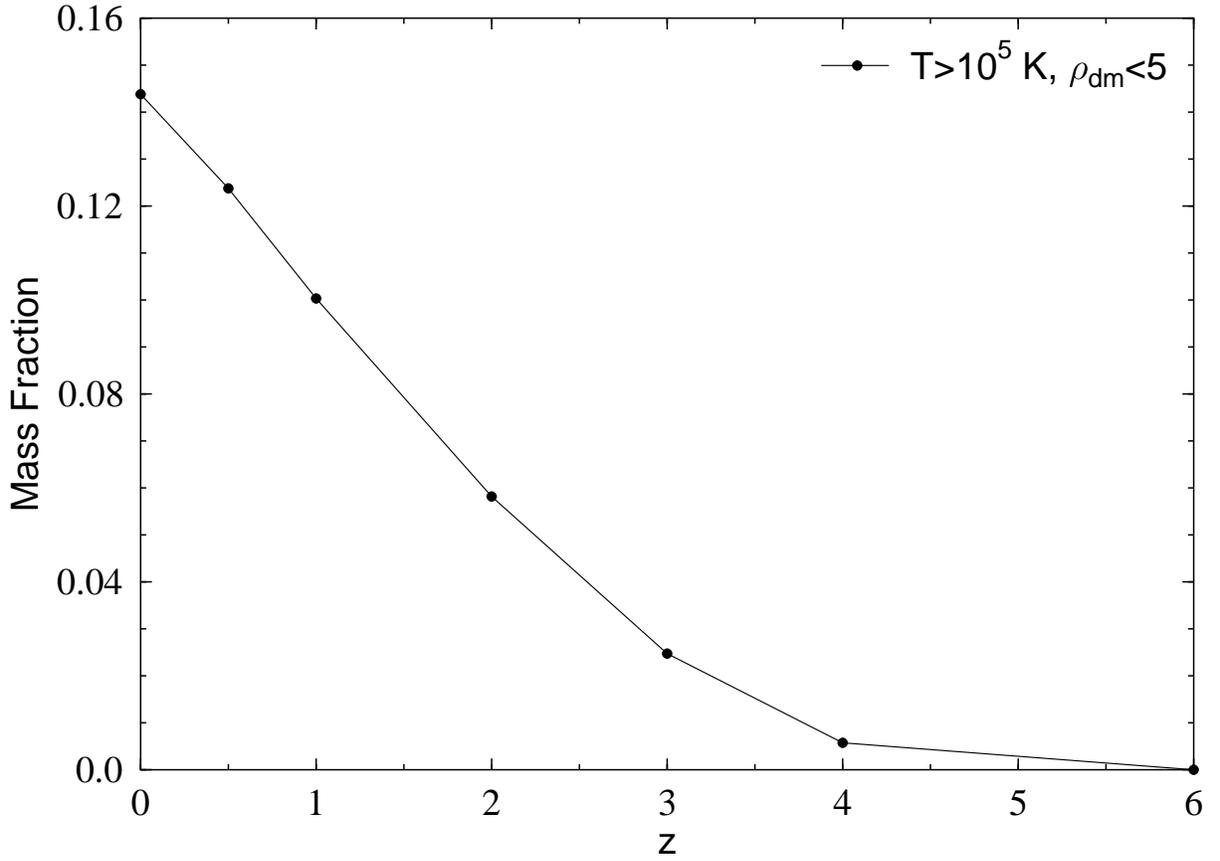}}

\caption{The $z$-evolution for the mass fraction of the HBFP
         (high baryon fraction phase) in the total baryonic
         matter.}
\end{figure*}

\newpage

\begin{table}[t]
\caption{Mean $F_b$ for Mass Density $\rho_{DM}$ and Redshift $z$}
\bigskip
\begin{tabular}{lllllll}
\tableline $\rho_{DM}$ ($\bar{\rho}_{DM}$) & $z$=0 &
z=0.5 &  z=1 &  z=2   &  z=3   &  z=4 \\
\tableline
0.03 -- 0.5      & 1.23 & 1.20 & 1.17 & 1.12 & 1.09 & 1.06 \\
0.5 -- 5.0       & 1.17 & 1.11 & 1.07 & 1.05 & 1.04 & 1.03 \\
5.0 -- 100       & 1.11 & 1.00 & 0.99 & 0.96 & 0.92 & 0.89 \\
100 -- $\infty$  & 0.80 & 0.89 & 0.81 & 0.83 & 0.72 &  N/A$^a$ \\
\tableline \multicolumn{7}{l} {$^a$Data are not available for
$\rho_{DM}>100$ at $z=4$.}
\end{tabular}
\end{table}

\end{document}